\documentclass[aip,apl,twocolumn,groupedaddress]{revtex4}
\input epsf
\usepackage{amsmath}

\begin{document}

\title {Tip induced doping effects in local tunnel spectra of graphene}
\author{Shyam K. Choudhury and Anjan K. Gupta}
\affiliation{Department of Physics, Indian Institute of Technology Kanpur, Kanpur 208016, India.}
\date{\today}

\begin{abstract}
We report on tip induced doping in local tunnel spectra of single layer graphene (SLG) with tunable back-gate using room temperature scanning tunneling microscopy and spectroscopy (STM/S). The SLG samples, prepared on silicon dioxide surface by exfoliation method and verified by Raman spectra, show atomically resolved honeycomb lattice. Local tunnel spectra show two minima with a clear evolution in the position of both with doping by the back gate. A similar variation in spectra is also observed spatially due to charge inhomogeneity. With doping the two minima move in opposite directions with one showing nearly a square root dependence and the other a linear dependence on gate voltage. We explain these features as arising from the STM tip induced and bias voltage dependent doping in SLG.
\end{abstract}

\maketitle

The electrical transport in graphene and its dependence on carrier concentration has revealed some spectacular physics in this Dirac-like two dimensional electronic system \cite{rev}. The carrier concentration in single layer graphene (SLG) has mainly been controlled by a silicon back gate separated from graphene by a 300nm thick SiO$_2$ layer. Two gates, one on top in addition to the bottom gate, have been used to create a tunable gap in the bilayer graphene \cite{bil-gap}. Recently, effects of local doping on bulk transport have also been reported using scanning gate microscopy \cite{scan-gate} and scanning tunneling microscopy (STM) \cite{stm-ucf} which use the microscope tip to induce doping. Scanning tunneling microscopy and spectroscopy (STM/S) is an ideal tool for locally probing the tip induced doping.
\begin{figure}[tbp]
\centerline{\epsfysize = 2.6 in \epsfbox{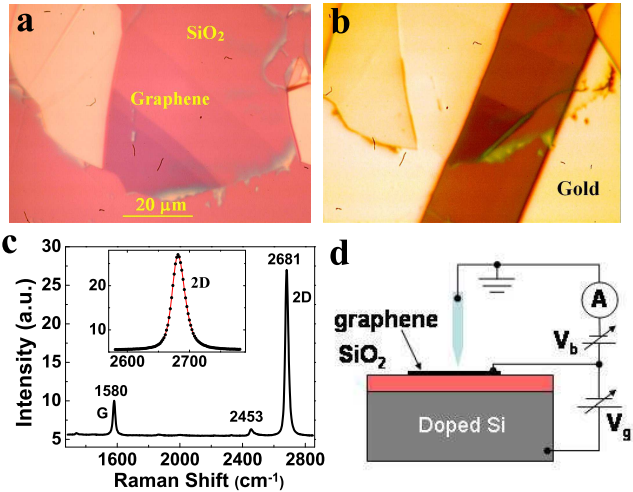}}
\caption{\textbf{a.} Optical image of few layers graphene (FLG) with a portion (marked as 'Graphene') of SLG on SiO$_2$ \textbf{b.}
The FLG flake after deposition of Cr/Au contact leads. \textbf{c.} Raman Spectra of SLG region in ambient conditions using laser light of wavelength 514 nm, spot size 1$\mu$m and power 0.64 mW. The inset shows the details of the 2D peak centered at 2681 cm$^{-1}$. \textbf{d.} STM measurement schematic of this graphene device.}
\label{fig:optical-raman}
\end{figure}
Here we report on tip induced doping effects in local tunnel spectra on SLG prepared by mechanical exfoliation on SiO$_2$/Si surface. Local tunnel spectra exhibit two minima with the two moving in opposite directions along the bias axis with increasing gate voltage. We explain the two minima and their evolution with gate voltage in terms of the bias voltage dependent doping by the STM tip.

\begin{figure}[htbp]
\centerline{\epsfysize=3.0in \epsfbox{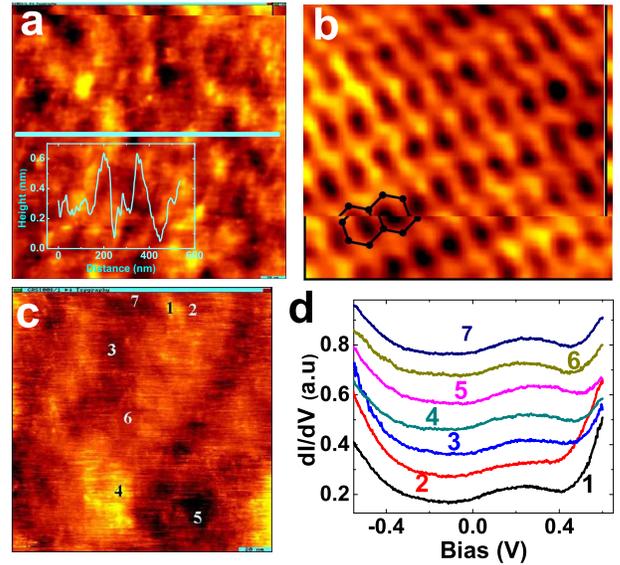}}
\caption{\textbf{a.} Topographic image of SLG (560$\times$560 nm$^2$ at 0.4V and 0.1nA). The inset shows topographic profile along the white line shown in the image \textbf{b.} Atomically resolved constant-current STM image (2.4$\times$2.1 nm$^2$ at 0.4 V) of SLG. \textbf{c.} Topographic image of SLG (161$\times$161 nm$^2$ at 0.4V and 0.1nA). \textbf{d.} Tunneling conductance at various positions shown in \textbf{c}.}
\label{fig:topography}
\end{figure}
The SLG samples were prepared on 300 nm thick insulating SiO$_2$ film on doped silicon substrate by mechanical exfoliation of graphite flakes \cite{graphite}.  SLG as a part of few layers graphene (FLG), size($\approx$ 50$\mu$m $\times$ 18$\mu$m) was optically identified \cite{optical-detection} as shown in Fig.\ref{fig:optical-raman}a. For connecting leads we deposited $\sim$1 mm wide and few mm long strips of 10nm Cr topped by 100 nm Au by masking larger portion of graphene using a tungsten wire of 25 $\mu$m diameter (see Fig.\ref{fig:optical-raman}b). Such mechanical masking \cite{mech-mask} minimizes the contamination from the liquid chemicals used in lithographic techniques. The $\mu$-Raman spectrum in Fig.\ref{fig:optical-raman} shows the G peak at 1580 cm$^{-1}$, 2D peak at 2681 cm$^{-1}$ and the minor D peak at 1339 cm$^{-1}$. This confirms this region to be SLG \cite{raman-rivew}.

The STM data reported here were taken at room temperature using a homemade STM in a cryo-pumped vacuum chamber with 10$^{-4}$ mbar pressure. The gate voltage ($V_g$) was applied to the silicon substrate with 270 k$\Omega$ series resistance. For STM measurements, as depicted in the schematic in Fig. \ref{fig:optical-raman}d, we apply the bias voltage ($V_b$) to the sample while the tip stays at (virtual) ground potential. A homemade 2D nano positioner \cite{nano-positioner} was used for aligning SLG with the STM tip under an optical microscope. Electrochemically etched and HF treated \cite{hf-cleaning} tungsten wire of 0.25 mm diameter was used as the STM tip. The apex radius of the tip was found to be in 20-50nm range from electron microscopy. The tunnel spectra were acquired using an ac modulation of amplitude 20 mV with the dc bias voltage. The junction resistance was kept as 4G$\Omega$ (400mV bias and 100pA tunnel current) for all the local spectra. We have checked reproducibility of the spectra with several tungsten tips on different SLG samples but the data presented here are from the sample shown in Fig.\ref{fig:optical-raman}.

Fig.\ref{fig:topography}a shows the topographic image of a large area (560$\times$560 nm$^2$) of the SLG region of the sample. The yellow and dark regions are bumps and troughs either due to charge inhomogeneity or ripples as observed in previous STM studies \cite{combo-stm-slg,loc-dop}. Inset of this image also shows the topographic profile, with a roughness of $\sim$0.2nm, along the white line marked in the image of Fig.\ref{fig:topography}a. Fig.\ref{fig:topography}b is the atomic resolution image showing the honeycomb lattice structure of SLG. Fig.\ref{fig:topography}c shows a topographic image of area 161$\times$161 nm$^2$ with the tunnel spectra at the marked locations shown in the Fig.\ref{fig:topography}d. These spectra are taken at zero gate voltage ($V_g$). All of the spectra show two minima contrary to the expected `V' shape DOS for SLG. The first minima (closer to zero bias) has less curvature than the second minima. The position of the second minima on the bias axis varies between 0.4 and 0.5 V at different locations. The location of the two minima varies as one moves to different locations of the sample due to local charge density \cite{loc-dop} as we discuss later.

We can understand the two minima if we incorporate the doping arising from the tip bias voltage as follows. Using the linear dispersion for graphene near Dirac point \cite{rev}, i.e. $\epsilon_{{\bf k} \pm}= \pm\hbar v_F |\bf k|$ with $v_F=10^6$ m/s as the Fermi velocity, and the 2-fold spin and 2-fold valley degeneracy, the DOS of SLG is given by,
\begin{eqnarray}
N(E) = \frac{2}{\pi \hbar^2 {v_F}^2} \left|E+E_{F}\right|.
\label{eq:dos}
\end{eqnarray}
\begin{figure}[htbp]
\centerline{\epsfxsize=3.6in \epsfbox{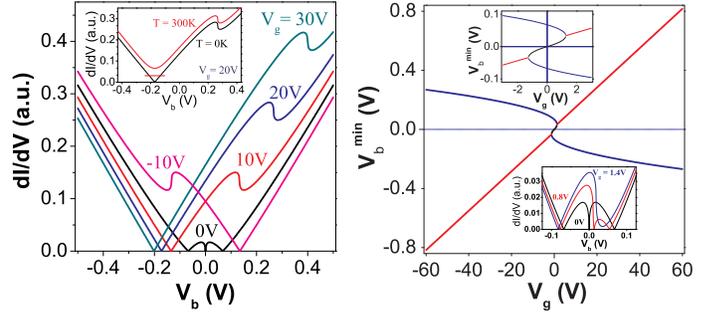}}
\caption{\textbf{a.} Theoretical tunneling conductance at $T = 0$K for different gate voltages. The inset shows a comparison between $T= 0$K and 300K for $V_g=20$V. \textbf{b.} Variation of primary and secondary minima with $V_g$. The bottom inset shows the low bias region of the spectra at small $V_g$ values showing the fine spectral features. The top inset shows the minima locations for small $V_g$ and $V_b$ values. All the plots are for $\beta=75$ and $\gamma=0.032$V$^{1/2}$.}
\label{fig:spectra-theory}
\end{figure}
Here, $E_F$ is the Fermi energy measured from the Dirac point, and determined by the carrier density, and $E$ is the energy measured with respect to $E_F$. For a given number density of electrons, $n$, and using Eq. \ref{eq:dos} we find $E_F = \hbar v_F n \sqrt{\pi/|n|}$. Assuming a parallel plate capacitor configuration (see Fig. \ref{fig:optical-raman}d) of SLG with both the bottom gate electrode and the STM tip \cite{cap-just}, we can write,
\begin{eqnarray}
n=\frac {K \epsilon_0} {ed} (V_{g}-\beta V_{b}).
\label{eq:cdensity}
\end{eqnarray}
Here, $\beta=\frac{d}{z K}$ with $d$ as the oxide layer thickness, $z$ is the tip-sample separation and $K$ ($\approx4$) is the dielectric constant of SiO$_2$. $e$ is the magnitude of electronic charge, $\epsilon_0$ is permeability of free space, $V_{g}$ is the gate voltage and $V_{b}$ is the bias voltage applied to the sample keeping the STM tip at zero potential. From here we see that although the magnitude of $V_b$ is smaller, by about two orders of magnitude, than $V_g$, the factor $\beta$ (= 75 for $z=1$nm) makes the two terms comparable to each other. Thus the tip induced doping effects should be easily observable.

Using Eq. \ref{eq:cdensity} for $n$ to find $E_F$, which in turn can be used in Eq. \ref{eq:dos} to get DOS as, $N(E)= \frac{2}{\pi\hbar^2{v_F}^2}\left| E+\gamma e \frac{V_{g}-\beta V_{b}}{\sqrt{|V_{g}-\beta V_{b}|}}\right|$. Here, $\gamma=\hbar v_F\sqrt{\frac{K\epsilon_{0} \pi}{e^3 d}} =0.032$ V$^{1/2}$ for $v_F=10^6$ m/s, $d=300nm$ and $K=4$. Thus the tunnel conductance at zero temperature is given by
\begin{eqnarray}
G(V_b,T=0)=e\alpha \left|{V_b+\frac{\gamma (V_{g}-\beta V_{b})}{\sqrt{|V_{g}-\beta V_{b}|}}}\right|.
\label{eq:tun-cond}
\end{eqnarray}
Here is $\alpha$ is a constant. Since the tunnel conductance at finite temperatures is proportional to the convolution of DOS and negative of the derivative of the Fermi function, $f(E,T)$, \cite{chen}, we get, $G(V_b,T)=\alpha \int^\infty_{-\infty} \left|{E+\frac{\gamma e(V_{g}-\beta V_{b})}{\sqrt{|V_{g}-\beta V_{b}|}}}\right|\left(-\frac{\partial f(E-eV_b,T)}{\partial E}\right)dE$. Fig. \ref{fig:spectra-theory}a shows the calculated spectra for $\beta=75$ and $\gamma=0.032$ V$^{1/2}$ with a minima other than the Dirac point and with a clear evolution with $V_g$. At room temperature the Dirac point also appears as a minima due to thermal smearing and local disorder. These two minima are consistent with the experimental spectra.
\begin{figure}[tbp]
\centerline{\epsfxsize = 3.6 in \epsfbox{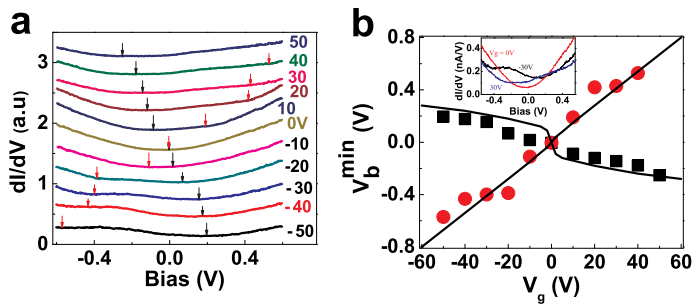}}
\caption{Experimental spectra at different $V_g$ values taken at a particular location on SLG. The spectra have been offset vertically uniformly for clarity. \textbf{b.}  Behavior of the two minima with $V_g$. Scattered data show experimental observation. Solid and dashed lines show theoretical expectation with $\beta=75$ and $\gamma=0.032$V$^{1/2}$. The inset shows three spectra (at $V_g=$+30, 0 and -30 V) without offset.}
\label{fig:min}
\end{figure}

Let us analyze the tunnel spectra features at $T=0$, given by Eq. \ref{eq:tun-cond}. Finite temperature only smears the spectra retaining the essential features. Due to square-root dependence of $E_F$ on $n$, $E_F$ has a sharp jump with $n$ near $n=0$, i.e. as $E_F$ passes through the Dirac point. This happens at $V_g=\beta V_b$ and gives rise to a minima in $G(V_b,0)$ at $|V_b|$ close to but slightly larger than $|V_g/\beta|$. The exact location of minimum at $T=0$ is found to be $V_b^{min}=\frac{V_g}{\beta}+\frac{\gamma^2\beta V_g}{4|V_g|}$ for $V_g\geq\frac{\beta^2 \gamma^2}{4}$ = 1.27V for $\beta=75$ and $\gamma=0.032$ V$^{1/2}$. Other than these minima the Dirac point exists at $V_b=-\frac{V_g}{2|V_g|}(\beta \gamma^2+\gamma \sqrt{4|V_g|+\beta^2 \gamma^2})$. There is more structure in the conductance spectra for $|V_g|<\frac{\beta^2 \gamma^2}{4}$ and at low bias ($|V_b|<\frac{\gamma^2\beta}{2}=38$ mV); however, it is not possible to observe these low bias features at room temperature due to limited energy resolution. Moreover, the low bias region may be significantly affected by disorder \cite{slg-dis-dos} and electron-phonon interactions \cite{slg-e-ph-int}. The spectra on our sample do not show a sharp cusp corresponding to Dirac point due to significant disorder as seen from the STS studies and also due to finite temperature. It is well known that SLG on SiO$_2$ has significant charge inhomogeneity due to local dopants that shifts the $E_F$ \cite{loc-dop} away from the Dirac point. In addition local dopants can also give rise to changes in DOS \cite{slg-imp-dos}. In general, some feature in the spectra should be observable whenever $E_F$ passes through a sharp feature in DOS.

We also note that the second minima, when far from the Dirac minima, does not get thermally smeared by temperature as seen in the inset of Fig.\ref{fig:spectra-theory}a. This happens as there is no actual minima in the DOS at this energy. This minimum comes from the linear dispersion at Dirac point and will occur only for SLG and thus the second minima will not be present in spectra taken on more than one layer of graphene though the location of the first minima may be affected by the tip induced doping.

Fig.\ref{fig:min}a shows the dependence of the spectra on gate controlled doping. These spectra were acquired at the same point of an image (not shown). We see two minima in these spectra with a clear evolution with the gate voltage. With increase of carrier concentration of either type the two minima move in opposite directions as expected. The exact locations of the minima were obtained by fitting a quadratic function in a small bias range ($\approx$100mV) around the minima and are indicated by the arrows. Fig.\ref{fig:min}b shows the comparison between the observed location of the two minima and the calculation with $z=1$nm. We see a good agreement between experiment and theory. Slight deviation of the minima from the expected behavior may arise from changes in spectra due to disorder \cite{slg-dis-dos} and impurities \cite{slg-imp-dos}.

\begin{figure}[htbp]
\centerline{\epsfxsize=3.6in \epsfbox{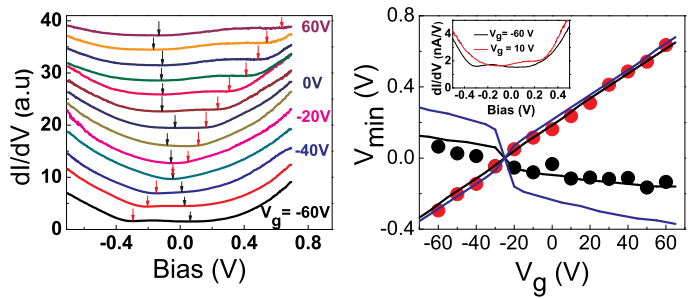}}
\caption{\textbf{a.} \textbf{a.} Spectra at different $V_g$ values (10V interval) taken at another location of SLG. The spectra have been offset vertically uniformly for clarity. \textbf{b.} Behavior of the two minima with $V_g$ with the scattered data showing experimental observations and solid line showing the calculated minia with $z=0.55$nm and $v_F=10^6$m/s. The dashed line shows the calculated minima $z=0.55$nm and $v_F=5\times10^5$m/s. The inset shows two spectra from \textbf{a} at gate voltages 10 and -60 V.}
\label{fig:spectra1}
\end{figure}

As discussed earlier, the two minima have significant spatial inhomogeneity arising from variations in local doping. Fig. \ref{fig:spectra1} shows the $V_g$ dependence of the two minima at a different location showing this region to be electron doped with a carrier density equivalent to $V_g\approx-25$V. In this case we obtain a good agreement with theory for the second minima using a tip-sample separation of 0.55 nm and incorporate the effect of electron doping due to charge impurity. The observed shift in primary minima is smaller as compared to the theory. We can fit both the minima, as shown in Fig. \ref{fig:spectra1}, by using a 50\% reduction in $v_F$ which may arise from the disorder \cite{biro} as this SLG portion is close to dopants causing disorder.

In conclusion, we have done a STM study of gated single layer graphene films prepared by mechanical exfoliation. The single layer graphene shows a topography of 0.3 nm over the 500 nm length scale. STS measurement at room temperature shows two minima whose shape and position changes with doping. We explain these features in detail on the basis of tip induced doping effects.

AKG acknowledges financial support from Department of Science and Technology and S. K. Choudhary acknowledges financial support from the University Grant Commission of the Government of India.


\begin{thebibliography}{99}
\bibitem{cap-just} From tip's electron microscopy images we find the radius of tip apex to be $\sim30$nm while the tunneling is rather local as confirmed from the atomic resolution images. This somewhat justifies the parallel plate capacitor approximation. In any case, what matters here is the linear relation between carrier density and the bias voltage which will be true for any capacitor model but with a different linear coefficient. The latter will depend on the tip-sample separation whose absolute value cannot be found with great precision in STM.
\end{thebibliography}
\end{document}